\documentclass[useAMS,usenatbib]{mn2e}

\usepackage{amsmath}
\usepackage{amsfonts}
\usepackage{amssymb}
\usepackage{epsf}
\usepackage{epsfig}
\usepackage{graphicx}

\title [Dark Energy in a Hyperbolic Universe]
{Dark Energy in a Hyperbolic Universe}
\author[R.~Aurich and F.~Steiner] {R.~Aurich and F.~Steiner\\
Abteilung Theoretische Physik, Universit\"at Ulm\\
Albert-Einstein-Allee 11, D-89069 Ulm \\ Germany }

\pubyear{2002} \volume{000} \pagerange{1}
\begin{document}

\maketitle \label{firstpage}

\begin{abstract}
Dark energy models due to a slowly evolving scalar (quintessence) field
$\phi$ are studied for various potentials $V(\phi)$ in a universe with
negative curvature.
The potentials differ in whether they possess a minimum at $\phi=0$
or are monotonically declining, i.\,e.\ have a minimum at infinity.
The angular power spectrum $C_l$ of the cosmic microwave background (CMB)
as well as the magnitude-redshift relation $m_{\text{B}}(z)$
of the type Ia supernovae are compared with the quintessence models.
It is demonstrated that some of the models with
$\Omega_{\text{tot}}\simeq 0.85\dots 0.9$ are in agreement with the
observations, i.\,e.\ possess acoustic peaks in the angular
power spectrum of the CMB anisotropy as observed experimentally,
and yield a magnitude-redshift relation in agreement with the data.
Furthermore, it is found that the power spectrum $P(k)$ of the
large-scale structure (LSS) agrees with the observations too.
\end{abstract}

\begin{keywords} 
cosmology:theory -- cosmic microwave background -- large--scale
structure of universe -- dark matter -- cosmological constant --
quintessence 
\end{keywords}

%%%%%%%%%%%%%%%%%%%%%%%%%%%%%%%%%%%%%%%%%%%%%%%%%%%%%%%%%%%%%%%%%%%%%%%%%%%%%

\section{Introduction}

The best experimental data constraining cosmological models is provided
by the observed cosmic microwave background (CMB) anisotropy
\citep{Netterfield_et_al_2001,Lee_et_al_2001,Halverson_et_al_2001}
and the magnitude-redshift relation of the type Ia supernovae
\citep{Hamuy_et_al_1996,Riess_et_al_1998,Perlmutter_et_al_1999}.
A further observational constraint is provided by the
power spectrum of the large-scale structure (LSS)
\citep{Peacock_Dodds_1994,Hamilton_Tegmark_Padmanabhan_2000,%
Percival_et_al_2001},
where there are now signs of oscillatory modulations
\citep{Percival_et_al_2001}.
The magnitude-redshift relation yields a strong hint that the
expansion of the universe is accelerating
which in turn is explained by a significant contribution of dark energy
to the energy content of the universe \citep{Perlmutter_et_al_1998}.
A long known candidate for dark energy is the vacuum energy leading
to the cosmological constant $\Lambda$ in the Einstein field equations.
An alternative is that the dark energy,
also called quintessence in this case,
is due to a slowly evolving scalar field $\phi$ with an
equation of state $-1 < w_\phi < 0$,
where $w_\phi=p_\phi/\varepsilon_\phi$ denotes the ratio of
pressure $p_\phi$ to energy density $\varepsilon_\phi$
\citep{Ratra_Peebles_1988,Peebles_Ratra_1988,%
Wetterich_1988_a,Wetterich_1988_b,Caldwell_Dave_Steinhardt_1998}.

Since inflationary models predict a flat universe,
dark energy scenarios are usually studied in flat models.
Currently favoured models assume a matter contribution of
$\Omega_{\text{m}}\simeq 0.3$ and a vacuum energy contribution of
$\Omega_\Lambda\simeq 0.7$.
Such models are, however, only marginally consistent
\citep{Boughn_Crittenden_2001}
with the cross-correlation of the CMB to the distribution of radio
sources of the NRAO VLA Sky Survey \citep{Condon_et_al_1998}.
A smaller contribution of vacuum energy would yield better
agreement with this data.
Since $\Omega_{\text{m}}$ is constrained by dynamical mass determinations,
this points to a universe with $\Omega_{\text{tot}}< 1$.
A further observation, which points in the same direction,
is provided by the statistics of strong gravitational lensing.
Too few lensed pairs with wide angular separation are observed
in comparison with the prediction of the currently favoured flat
$\Lambda$CDM-model \citep{Sarbu_Rusin_Ma_2001}.

In this paper we want to investigate how negative the spatial curvature
can be, i.\,e.\ how small can $\Omega_{\text{tot}}< 1$ be
in order to be in agreement with the observations?
The standard cosmological model based on the
Fried\-mann-Le\-ma\^{\i}tre-Robertson-Walker metric $(c=1)$
$$
ds^2 \; = \; a^2(\eta) \left\{ d\eta^2 - \gamma_{ij} dx^i dx^j
\right\}
\hspace{10pt} ,
$$
where $\gamma_{ij}$ denotes the spatial metric,
is governed for negative curvature ($K=-1$) by the Friedmann equation
\begin{equation}
\label{Eq:Friedmann}
{a'}^2 - a^2 \; = \; \frac{8\pi G}3 T_0^0 a^4
\hspace{10pt} ,
\end{equation}
where $a(\eta)$ is the cosmic scale factor and $\eta$ the conformal time.
Here and in the following,
the prime denotes differentiation with respect to $\eta$.
The energy momentum tensor $T_\mu^\nu$ contains the usual contributions
from relativistic components,
i.\,e.\ photons and three massless neutrino families,
non-relativistic components, i.\,e.\ baryonic and dark matter,
as well as the contribution arising from the scalar field $\phi$.
The energy density $\varepsilon_\phi$ of the scalar field is determined by
\begin{equation}
\label{Eq:energy_phi}
\varepsilon_\phi \; = \; \frac{1}{2 a^2} \, {\phi'}^2 + V(\phi)
\hspace{10pt} ,
\end{equation}
and the pressure $p_\phi$ is given by
\begin{equation}
\label{Eq:pressure_phi}
p_\phi \; = \; \frac{1}{2 a^2} \, {\phi'}^2 - V(\phi)
\hspace{10pt} ,
\end{equation}
where $V(\phi)$ denotes the quintessence potential.
Here it is seen that for a negative pressure $p_\phi$ a
slowly evolving scalar field is required.
Assuming that the scalar field couples to matter only through gravitation,
the equation of motion of the scalar field is
\begin{equation}
\label{Eq:equation_of_motion_phi}
\phi'' \, + \, 2 \frac{a'}{a} \phi' \, + \,
a^2 \frac{\partial V(\phi)}{\partial \phi} \; = \; 0
\hspace{10pt} .
\end{equation}
The Friedmann equation (\ref{Eq:Friedmann}) and
the continuity equation for the components $x$
\begin{equation}
\label{Eq:continuity_equation}
\varepsilon_x ' \, + \, 3 (1+w_x) \, \frac{a'}a \, \varepsilon_x \; = \; 0
\hspace{10pt} ,
\end{equation}
with $w_x=\frac 13$ for relativistic and $w_x=0$ for
non-relativistic components, determine together with
(\ref{Eq:equation_of_motion_phi}) the background model
and thus $\Omega_\phi(\eta)$.
The big-bang nucleosynthesis (BBN) provides a tight constraint on
the magnitude of $\Omega_\phi$ at the time $\eta_{\text{BBN}}$
of nucleosynthesis, which is $\Omega_\phi(\eta_{\text{BBN}}) < 0.13$
\citep{Ferreira_Joyce_1998}.
A stronger constraint $\Omega_\phi(\eta_{\text{BBN}}) < 0.045$
\citep{Bean_Hansen_Melchiorri_2001}
is obtained using new measurements of deuterium abundances.
All dark energy models discussed below have
$\Omega_\phi(\eta_{\text{BBN}}) \ll 10^{-5}$
and thus do not interfere with the standard BBN.
This is due to the fact that all considered potentials 
together with the chosen initial conditions maintain
negative $w_\phi$'s even well beyond the recombination epoch,
and thus $\Omega_\phi(\eta_{\text{BBN}})$ is negligible.
They are either no tracker potentials or the radiation dominated epoch
is too short to establish a tracker solution with a positive $w_\phi$.
The background model, which also yields the magnitude-redshift relation,
is further constrained by the observed Ia supernovae.

In our simulations we use the following parameters.
The Hubble constant
$H_0 = h \times 100 \hbox{km } \hbox{s}^{-1} \hbox{Mpc}^{-1}$
is set to $h=0.65$ lying close to the
average value of various determinations of the Hubble constant
\citep{Krauss_2001b}.
The density of baryonic matter $\Omega_{\text{b}}$ is set to
$\Omega_{\text{b}} = 0.05$ which leads to a value
$\Omega_{\text{b}} h^2 \simeq 0.021$ consistent with the
big-bang nucleosynthesis \citep{Tytler_et_al_2000} 
and to the DASI \citep{Pryke_et_al_2001} and
BOOMERanG \citep{Netterfield_et_al_2001} measurements.
The density of cold dark matter $\Omega_{\text{cdm}}$ is assumed
to be $\Omega_{\text{cdm}} = 0.35$.
Since we discuss models with negative curvature here,
the above values give $\Omega_\phi(\eta_0)<0.6$,
where $\eta_0$ is the present conformal time.

Among the various potentials discussed in the literature
(see e.\,g.\ \citet{Brax_Martin_Riazuelo_2000})
we concentrate on six different potentials.
Two examples of potentials $V(\phi)$ possessing a minimum at $\phi=0$
are the {\it cosine potential}
\citep{Frieman_Hill_Stebbins_Waga_1995,Coble_Dodelson_Frieman_1997,%
Kawasaki_Moroi_Takahashi_2001}
\begin{eqnarray}
\label{Eq:Pot_cos}
V(\phi) & = &
A (1-\cos(B \phi))
\\ & = & \nonumber
\frac 12\, A B^2 \,\phi^2 - \frac 1{24}\,AB^4\,\phi^4+ O\left(\phi^6\right)
\hspace{10pt} ,
\end{eqnarray}
and, as a special case of the potentials discussed in
\citep{Chimento_Jakubi_1996,Barreiro_Copeland_Nunes_2000,Sahni_Wang_2000},
the {\it hyperbolic-cosine potential} 
\begin{eqnarray}
\label{Eq:Pot_cosh}
V(\phi) & = &
A (\cosh(B \phi)-1)
\\ & = & \nonumber
\frac 12\, A B^2 \,\phi^2 + \frac 1{24}\,AB^4\,\phi^4+ O\left(\phi^6\right)
\hspace{10pt} .
\end{eqnarray}
Here, the initial condition for the scalar field is
$\phi_{\text{in}} \neq 0$ such that $\phi$ evolves towards the
minimum and then (possibly) oscillates around it.

Examples of monotonic potentials $V(\phi)$ are given by the
simple {\it exponential potential} \citep{Ratra_Peebles_1988}
\begin{equation}
\label{Eq:potential_exp}
V(\phi) \; = \; A \, \exp(-B \phi)
\end{equation}
which has for $B > \sqrt{24\pi(w_{\text{bg}}+1)}/m_{\text{p}}$ (in the flat case)
the remarkable property of self-adjusting $w_\phi$ to the
background equation of state $w_{\text{bg}}$ (scaling solution)
\citep{Copeland_Liddle_Wands_1998,Ferreira_Joyce_1997},
and the {\it inverse power potential}
\begin{equation}
\label{Eq:potential_power}
V(\phi) \; = \; \frac{A}{(B\phi)^\gamma}
\hspace{10pt} , \hspace{10pt}
\gamma > 0
\hspace{10pt} ,
\end{equation}
which is a simple example of a tracker field potential
\citep{Zlatev_Wang_Steinhardt_1999,Steinhardt_Wang_Zlatev_1999}.
(Here $m_{\text{p}} = G^{-1/2}$ denotes the Planck mass $(\hbar=1)$.)
These potentials are also obtained by requiring a constant $w_\phi$
during a given evolution stage of the universe.
The inverse power potential (\ref{Eq:potential_power}) is obtained
by requiring $w_\phi=\hbox{const.}$
during the radiation epoch, and the exponential potential
(\ref{Eq:potential_exp}) is obtained by this requirement during
the quintessence epoch.

In the following we define the above potentials by the dimensionless
parameters $\alpha$ and $\beta$ which are related to $A$ and $B$  by
$$
A =: (\alpha\, \hbox{eV})^4
\hspace{10pt} , \hspace{10pt}
B =: \beta / m_{\text{p}}
\hspace{10pt} .
$$
The parameter $\alpha$ is of the order $10^{-3}$ such that
$A$ corresponds to the expected vacuum energy density
$\varepsilon_\Lambda = \Lambda/8\pi G \simeq (0.003\, \hbox{eV})^4$.
This energy scale is, compared to other scales like $m_{\text{p}}$ or
$m_{\text{electroweak}}$, many orders of magnitude smaller
which results in the well-known fine tuning problem.
In the case of the cosine potential (\ref{Eq:Pot_cos})
a possible solution is suggested by \citet{Frieman_Hill_Stebbins_Waga_1995}
in the framework of a low-energy effective Lagrangian
describing an ultralight pseudo Nambu-Goldstone boson
with $A=M^4$ and $B=1/f$, where $M$ is expected to be of the order of
a light neutrino mass and $f$ is the global symmetry breaking scale.

%%%%%%%%%%%%%%%%%%%%%%%%%%%%%%%%%%%%%%%%%%%%%%%%%%%%%%%%%%%%%%%%%%%%%%%%%%%%%
\begin{figure}
\begin{center}
\vspace*{110pt}\includegraphics[width=8cm]{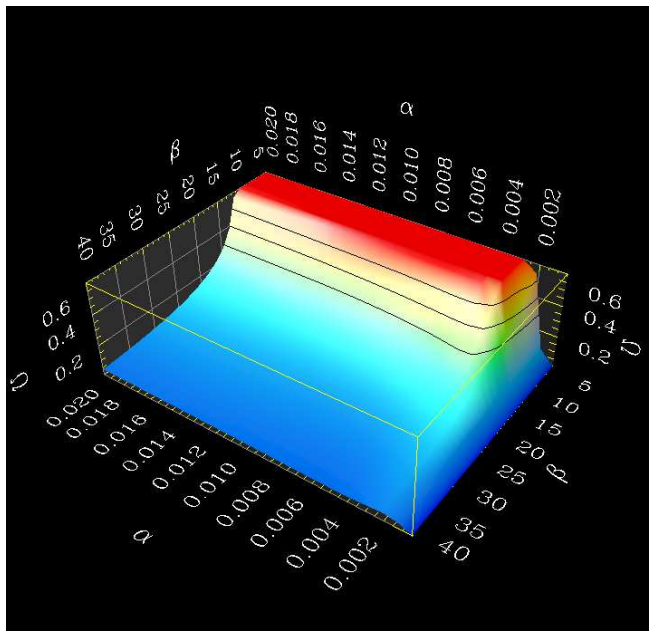}
\end{center}
\vspace*{-40pt}
\caption{\label{Fig:Omega_exp}
$\Omega=\Omega_\phi(\eta_0)$ is shown with dependence on $\alpha$ and $\beta$
for the exponential potential (\protect\ref{Eq:potential_exp})
and the initial condition $\phi_{\text{in}}=0$.
}
\end{figure}
%%%%%%%%%%%%%%%%%%%%%%%%%%%%%%%%%%%%%%%%%%%%%%%%%%%%%%%%%%%%%%%%%%%%%%%%%%%%%

The fine tuning can be considered with respect to the potential
parameters as well as with respect to the chosen initial value
$\phi_{\text{in}}$ of the scalar field.
However, this distinction cannot be sharply made since a change of
$\phi_{\text{in}}$ by $\Delta\phi_{\text{in}}$ can approximately be compensated
by a change in the potential parameters.
This is trivial in the case of the exponential potential
(\ref{Eq:potential_exp}) where such a change corresponds to a new
parameter $\widehat{A} = A \exp(-B\Delta\phi_{\text{in}})$.
Thus $\phi_{\text{in}}$ can be set to zero for the potential
(\ref{Eq:potential_exp}) without loss of generality.
In the case of the potentials (\ref{Eq:Pot_cos}), (\ref{Eq:Pot_cosh})
and (\ref{Eq:potential_power}) we use in the following the initial
condition $\phi_{\text{in}} \neq 0$.
The kinetic energy ${\phi'}^2/(2a^2) $ is for the potentials
(\ref{Eq:Pot_cos}) to (\ref{Eq:potential_power}) initially zero,
i.\,e.\ initially one has $w_\phi(0)=-1$, since it follows
from the Klein-Gordon equation (\ref{Eq:equation_of_motion_phi})
that $\phi'(\eta)$ vanishes like $O(\eta^3)$ in the
radiation-dominated epoch, $\eta\to 0$, assuming
$\frac{\partial V}{\partial \phi}(\phi_{\text{in}}) \neq 0$.

The above potentials depend on the parameters $\alpha$ and $\beta$,
respectively $\alpha$, $\beta$ and $\gamma$.
For a fixed $\Omega_\phi(\eta_0)$, where $\eta_0$ is determined by
the condition that the Hubble parameter of the model coincides with the
assumed Hubble constant $h=0.65$,
this gives a curve in the $\alpha$-$\beta$ parameter space
which, in general, is a one-parameter curve $\beta = \beta(\alpha)$.
However, there are exceptions for the potentials possessing a minimum
which can have multiple solutions for $\beta$ for fixed $\alpha$ and
$\Omega_\phi(\eta_0)$.
The inset of figure \ref{Fig:Cl_cosh_fit} shows such a behavior
for the potential (\ref{Eq:Pot_cosh}) near $\alpha = 0.0057$,
where three solutions exist for $\Omega_\phi(\eta_0)=0.5$.
The determination of $\beta(\alpha)$ depends also on $\phi_{\text{in}}$,
i.\,e.\ altering $\phi_{\text{in}}$ gives for a fixed
$\Omega_\phi(\eta_0)$ different values for $\alpha$ and $\beta$.
This is due to the fact that $\Omega_{\text{tot}}$ is not,
as in the flat case, constrained by the Friedmann equation itself.

The CMB anisotropy is computed according to \citet{Ma_Bertschinger_1995}
using the conformal Newtonian gauge.
As mentioned above, the relativistic components
are three massless neutrino families and photons
with standard thermal history.
For the photons, the polarization dependence on the Thomson cross section
is taken into account.
The recombination history of the universe is computed using RECFAST
\citep{Seager_Sasselov_Scott_1999}.
The non-relativistic components are baryonic and cold dark matter.
The initial conditions are given by an initial curvature perturbation
with no initial entropy perturbations in relativistic and
non-relativistic components.
Furthermore, we assume that there are no tensor mode contributions.
The initial curvature perturbation is assumed to be scale-invariant
which is ``naturally'' suggested by inflationary models.
The inhomogeneity of the scalar field is initially set to zero.
Other choices for this inhomogeneity,
e.\,g.\ setting it equal to the inhomogeneity of the matter component,
would yield the same results
because $\Omega_\phi$ is initially negligible.
The evolution of the scalar field inhomogeneity is computed
along the lines of \citet{Hu_1998}.
\citet{Hu_1998} develops a model for generalized dark matter (GDM),
where the GDM is characterized by the equation of state $w_{\text{GDM}}$,
the effective sound speed $c^2_{\text{eff}}$ and a
viscosity parameter $c^2_{\text{vis}}$.
This characterization of GDM includes radiation as well as cold dark matter.
The choice $c^2_{\text{eff}}=1$ and $c^2_{\text{vis}}=0$ gives
an exact description of a scalar field with $w_\phi=w_{\text{GDM}}$
(see Appendix in \citep{Hu_1998}) as studied in the present paper.
The equations (5) and (6) in \citep{Hu_1998} are used to describe
the evolution of the inhomogeneities of the scalar field.
Furthermore, no reionization is taken into account
because reionization occurs late around $z\sim 6$
as the recent detection of a Gunn-Peterson trough in a $z=6.28$
quasar shows \citep{Becker_et_al_2001}.
The CMB anisotropy computations also yield the linear power spectrum
$P(k)$ of the LSS.

The behavior of the scalar field $\phi$ is determined by the
damping term in (\ref{Eq:equation_of_motion_phi}),
i.\,e.\ whether the initial conditions and the potential are chosen such
that the field evolution is over-damped and the field is frozen to its
initial value leading to a model nearly indistinguishable from
a cosmological constant,
or whether the damping plays only a minor role and the field
evolves towards the potential minimum.
(See \citet{Coble_Dodelson_Frieman_1997,Kawasaki_Moroi_Takahashi_2001}
for a discussion in the case of the cosine potential.)
Figure \ref{Fig:Omega_exp} displays the dependence of $\Omega_\phi(\eta_0)$
on the parameters $\alpha$ and $\beta$ for the exponential potential
(\ref{Eq:potential_exp}) and the initial condition $\phi_{\text{in}}=0$.
Since we only discuss models with negative curvature, i.\,e.\ with
$\Omega_\phi(\eta_0)<0.6$, we show in figure \ref{Fig:Omega_exp}
the parameter surface for
$0 \leq \Omega_\phi(\eta_0) \leq 0.6$ only.
Thus the plateau seen in figure \ref{Fig:Omega_exp} at
$\Omega_\phi(\eta_0)=0.6$
corresponds to the cross section at $\Omega_{\text{tot}}=1$,
i.\,e.\ values of $(\alpha,\beta)$ above this plateau correspond
to solutions with positive curvature.
Also shown are three curves corresponding to 
$\Omega_\phi(\eta_0)=0.3, 0.4, 0.5$, respectively.
The over-damped domain corresponds to small values of $\beta$,
i.\,e.\ to $\alpha \sim 0.002$ in order to give a significant value
$\Omega_\phi(\eta_0)$ today.
For the potentials (\ref{Eq:Pot_cos}) and (\ref{Eq:Pot_cosh})
one obtains similar figures.
The potential (\ref{Eq:potential_power}), however, does not display
a sharp edge at which the evolution changes from frozen to rapidly
evolving.

\section{Potentials with a minimum}

%%%%%%%%%%%%%%%%%%%%%%%%%%%%%%%%%%%%%%%%%%%%%%%%%%%%%%%%%%%%%%%%%%%%%%%%%%%%%
\begin{figure}
\begin{center}
\begin{minipage}{4cm}
\vspace*{20pt}
\hspace*{-45pt}\includegraphics[width=3.8cm]{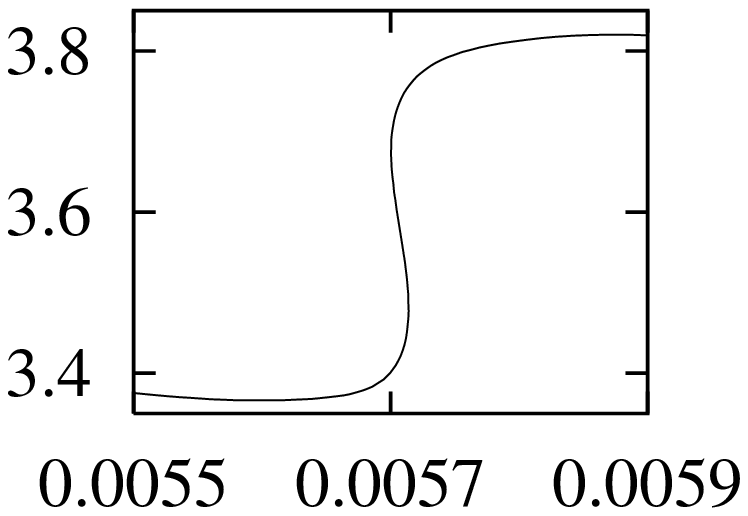}
\put(-34,6){$\alpha$}
\put(-90,50){$\beta$}
\end{minipage}
\begin{minipage}{9cm}
\vspace*{-100pt}\hspace*{-15pt}\includegraphics[width=9cm]{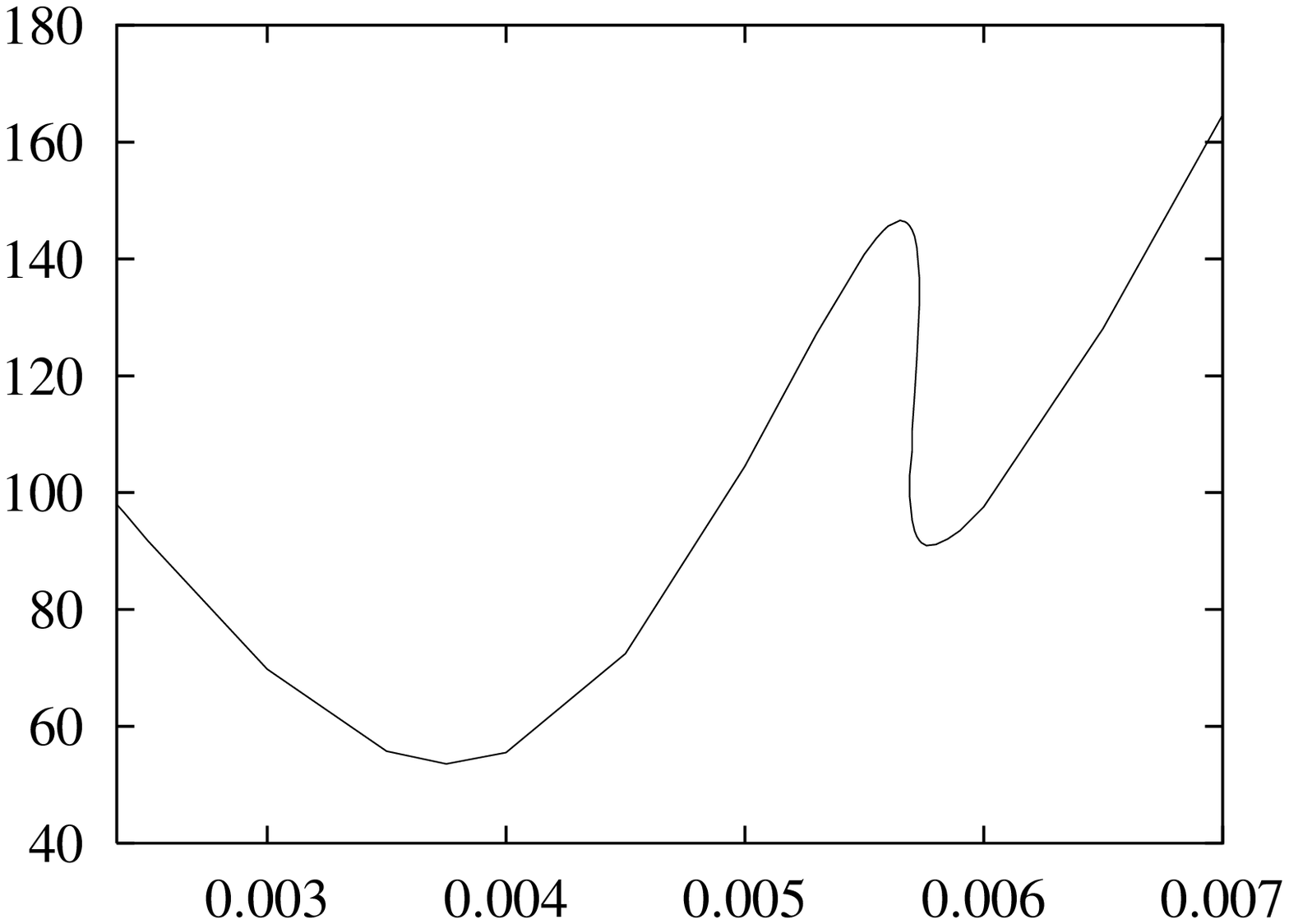}
\put(-42,7){$\alpha$}
\put(-237,158){$\chi^2$}
\end{minipage}
\end{center}
\vspace*{-10pt}
\caption{\label{Fig:Cl_cosh_fit}
The $\chi^2$ values of the fit to 41 data points of the BOOMERanG,
MAXIMA-1 and DASI experiments for the potential (\ref{Eq:Pot_cosh})
in dependence on the parameter $\alpha$.
The parameter $\beta$ is chosen such that $\Omega_\phi(\eta_0)=0.5$,
i.\,e.\ $\Omega_{\text{tot}} = 0.9$.
The inset shows the dependence of $\beta$ on $\alpha$ near
$\alpha=0.0057$.
}
\vspace*{-3pt}\end{figure}
%%%%%%%%%%%%%%%%%%%%%%%%%%%%%%%%%%%%%%%%%%%%%%%%%%%%%%%%%%%%%%%%%%%%%%%%%%%%%

\begin{figure}
\begin{center}
\hspace*{-25pt}\includegraphics[width=9cm]{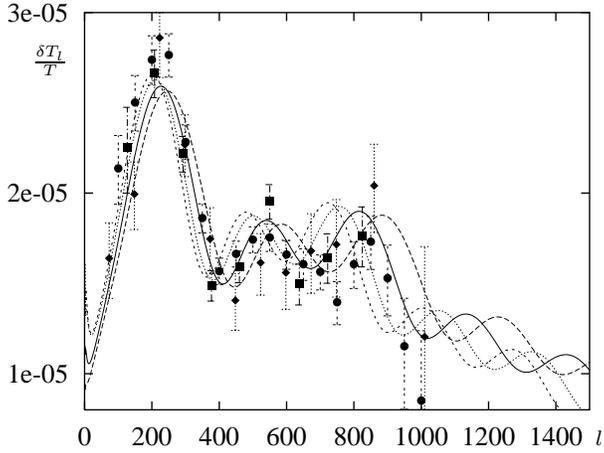}
\put(-15,7){$l$}
\put(-228,150){$\frac{\delta T_l}T$}
\end{center}
\vspace*{-15pt}
\caption{\label{Fig:Cl_cosh}
The CMB angular power spectrum $\frac{\delta T_l}T$ is shown for four
$\Omega_{\text{tot}} = 0.9$ models with the hyperbolic-cosine potential
(\protect\ref{Eq:Pot_cosh})
and $\alpha=0.00237$ (long dashed curve), $\alpha=0.00375$ (solid curve),
$\alpha=0.0056$ (short dashed curve), and $\alpha=0.0058$ (dotted curve).
Also shown is the experimental data from BOOMERanG (circles),
MAXIMA-1 (diamonds), and DASI (squares).
}
\end{figure}

We now discuss the dependence of the angular power spectrum
$\delta T_l=\sqrt{l(l+1)C_l/(2\pi)}$ of the CMB anisotropy
on the potentials (\ref{Eq:Pot_cos}) and (\ref{Eq:Pot_cosh}).
The $\delta T_l$ are compared with the 41 data points from
BOOMERanG \citep{Netterfield_et_al_2001},
MAXIMA-1 \citep{Lee_et_al_2001},
and DASI \citep{Halverson_et_al_2001}.
The amplitude of the initial curvature perturbation is fitted
such that the value of $\chi^2$ is minimized with respect to 
these three experiments,
where $\chi^2$ is computed using
RADPACK\footnote{See RADPACK homepage:\\
http://bubba.ucdavis.edu/\~\,knox/radpack.html}.
Figure \ref{Fig:Cl_cosh_fit} shows the dependence of $\chi^2$ on
the parameter $\alpha$ for the hyperbolic-cosine potential (\ref{Eq:Pot_cosh})
with $\Omega_\phi(\eta_0)=0.5$, i.\,e.\ $\Omega_{\text{tot}}=0.9$.
The frozen case around $\alpha=0.0024$, mimicking a cosmological constant,
gives a poorer fit than values around $\alpha=0.00375$.
With increasing $\alpha$ the fit deteriorates until a second
minimum is reached at $\alpha=0.0058$.
The rapid dependence of $\chi^2$ on $\alpha$ near $\alpha=0.0057$
is due to the form of the solution curve of $\Omega_\phi(\eta_0)=0.5$
in the $\alpha$-$\beta$ parameter space as shown by the inset.
Figure \ref{Fig:Cl_cosh} shows the $\delta T_l/T$ spectra for four different
models in comparison to the used experimental data.
The models belong to four different values of $\alpha$,
i.\,e.\ $\alpha=0.00237$, $\alpha=0.00375$, $\alpha=0.0056$,
and $\alpha=0.0058$, respectively.
These values were obtained for the initial condition
$\phi_{\text{in}}=1/B=m_{\text{p}}/\beta$.
As explained in the Introduction, this choice does not represent
a restriction since another choice for $\phi_{\text{in}}$ would
be possible as well, but then other parameter values would be obtained
in order to leave $\Omega_{\text{tot}}(\eta_0)$ at the assumed value.
This is in contrast to flat models where $\Omega_{\text{tot}}$ is
constraint to $\Omega_{\text{tot}}=1$ for all times by the
Friedmann equation.

%%%%%%%%%%%%%%%%%%%%%%%%%%%%%%%%%%%%%%%%%%%%%%%%%%%%%%%%%%%%%%%%%%%%%%%%%%%%%
%%%%%%%%%%%%%%%%%%%%%%%%%%%%%%%%%%%%%%%%%%%%%%%%%%%%%%%%%%%%%%%%%%%%%%%%%%%%%
%%%%%%%%%%%%%%%%%%%%%%%%%%%%%%%%%%%%%%%%%%%%%%%%%%%%%%%%%%%%%%%%%%%%%%%%%%%%%
\begin{figure}
\vspace*{-10pt}
\begin{center}
\hspace*{-10pt}\includegraphics[width=9cm]{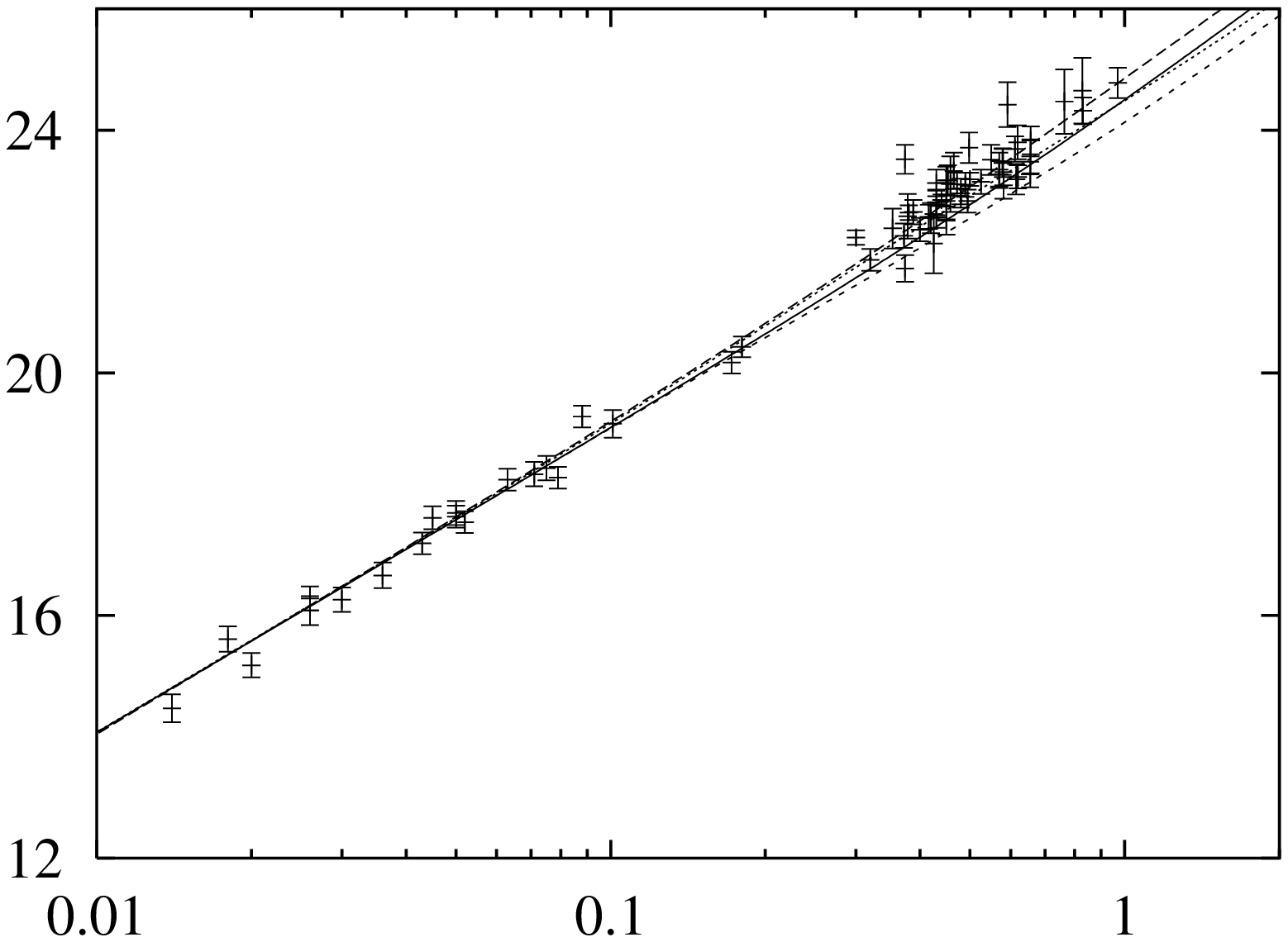}
\put(-25,7){$z$}
\put(-242,162){$m_{\text{B}}$}
\end{center}
\vspace*{-15pt}
\caption{\label{Fig:SN_cosh}
The magnitude-redshift relation $m_{\text{B}}(z)$ is shown for the same
models as in figure \ref{Fig:Cl_cosh} in comparison with the
supernovae Ia data.
}
\end{figure}
%%%%%%%%%%%%%%%%%%%%%%%%%%%%%%%%%%%%%%%%%%%%%%%%%%%%%%%%%%%%%%%%%%%%%%%%%%%%%
%%%%%%%%%%%%%%%%%%%%%%%%%%%%%%%%%%%%%%%%%%%%%%%%%%%%%%%%%%%%%%%%%%%%%%%%%%%%%
\begin{figure}
\vspace*{-10pt}
\begin{center}
\hspace*{-20pt}\includegraphics[width=9cm]{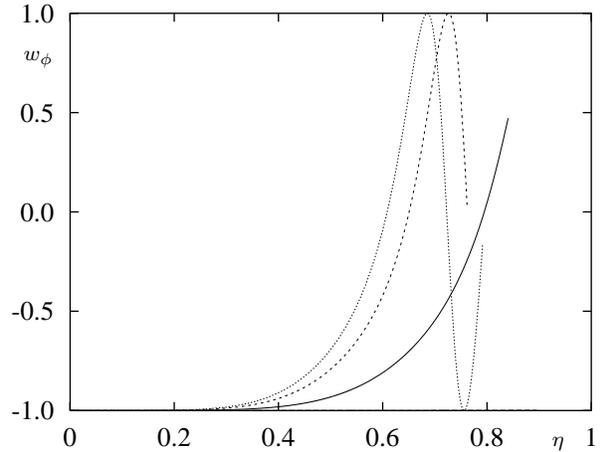}
\put(-32,7){$\eta$}
\put(-232,153){$w_\phi$}
\end{center}
\vspace*{-15pt}
\caption{\label{Fig:w_cosh}
The equation of state $w_\phi=p_\phi/\varepsilon_\phi$
is shown as a function of conformal time $\eta$
for the same models as in figures \ref{Fig:Cl_cosh}
and \ref{Fig:SN_cosh}.
The curves end at $\eta_0$, where the Hubble parameter of
the models coincides with the assumed Hubble constant.
}
\end{figure}
%%%%%%%%%%%%%%%%%%%%%%%%%%%%%%%%%%%%%%%%%%%%%%%%%%%%%%%%%%%%%%%%%%%%%%%%%%%%%

The location of the first peak depends mainly on the sound horizon
at last scattering and on the angular diameter distance to the
surface of last scattering (SLS).
For the models with the potential (\ref{Eq:Pot_cosh}),
shown in figure \ref{Fig:Cl_cosh},
the sound horizon is nearly the same
because of $\Omega_\phi(\eta_{\text{SLS}}) < 10^{-7}$.
Since the angular diameter distance becomes smaller for larger
values of $\alpha$, i.\,e.\ for a more dynamical field,
the location of the peaks is shifted towards lower multipoles $l$.
This general trend is slightly modified by the oscillatory behavior
of the field as is revealed by figure \ref{Fig:Cl_cosh}
where the first peak lies at $l=246$, $l=227$, $l=201$ and $l=210$
for $\alpha=0.00237$, $\alpha=0.00375$, $\alpha=0.0056$ and
$\alpha=0.0058$, respectively.
Thus larger values of $\alpha$ yield a first peak near $l\simeq 200$
even for $\Omega_{\text{tot}}<1$.
The position of the first peak for the model
with $\alpha=0.00237$ is at a value of $l$ which is too high.
For the models with larger $\alpha$ the positions of the first peak are
in agreement with the observations.
The second and third peaks for the models with
$\alpha=0.0056$ and $\alpha=0.0058$ occur at values of $l$ which are too low.
The best description of the data is therefore given by the model
with $\alpha=0.00375$.

Although the two models with the largest $\alpha$-values seem
to be similar with respect to their $C_l$ spectra,
their magnitude-redshift relation is different
as shown in figure \ref{Fig:SN_cosh}.
Here, the magnitude-redshift relation $m_{\text{B}}(z)$ is compared with
the data from \citet{Hamuy_et_al_1996,Riess_et_al_1998,Perlmutter_et_al_1999}
assuming an absolute magnitude of $M^B_{\text{MAX}}=-19.25$
of supernovae Ia \citep{Hamuy_et_al_1996}.
The data is well described by the models with $\alpha=0.00237$,
$\alpha=0.00375$ and $\alpha=0.0058$,
but the curve belonging to $\alpha=0.0056$ gives values which are too low.
Thus, the models with $\alpha=0.00375$ and $\alpha=0.0058$ are in agreement
with the positions of the first acoustic peaks and the supernovae observations.
The reason for this complementary behavior is revealed by figure
\ref{Fig:w_cosh}, where the evolution of the equation of state
$w_\phi$ as a function of conformal time $\eta$ is shown.
The model with $\alpha=0.00237$ possesses a nearly frozen field, and thus
$w_\phi$ increases only from $w_\phi(0)=-1$ to $w_\phi(\eta_0)\simeq -0.998$
which cannot be seen in figure \ref{Fig:w_cosh}.
The model with $\alpha=0.00375$ has $w_\phi(\eta_0)\simeq +0.5$
whereas the remaining two models with $\alpha=0.0056$ and $\alpha=0.0058$
have already passed a phase with $w_\phi=+1$, in which the
potential vanished and the kinetic term dominated.
The supernovae measurements test the more recent history up to $z=1$,
which corresponds to $\eta(z=1) \simeq 0.60$ for the latter two models.
Therefore, the model with $\alpha=0.0058$ possesses a more negative
effective $w_\phi$ averaged over $z$ from $z=1$ to $z=0$.
This in turn gives a better match with the supernovae observations.

The models have to provide an age of the universe compatible with
the mean age of the oldest globular clusters of
$12.7^{+3}_{-2} \hbox{Gyr}$ \citep{Krauss_2001b}.
The nearly frozen field case $\alpha=0.00237$ has an age of 13.0 Gyr,
the model with $\alpha=0.00375$ has an age of 11.6 Gyr,
whereas the model with $\alpha=0.0056$ has a low age of 9.9 Gyr
because of its large mean value of $w_\phi$.
The case with $\alpha=0.0058$ leading to a lower mean value of
$w_\phi$ gives 10.9 Gyr.
In the last two cases the age may be too low if the error bounds
in the age determination of globular clusters tighten.
Altogether we see that the model with $\alpha=0.00375$
describes the CMB and supernovae data quite well and yields
an age of the universe consistent with the oldest globular clusters.

Using the above initial conditions for the cosine potential
(\ref{Eq:Pot_cos}) gives very similar results.
The main difference is observed in the angular power spectrum
for small values of $l$.
The increase in power for $l \lesssim 10$ in figure \ref{Fig:Cl_cosh}
is due to an increased integrated Sachs-Wolfe contribution
because of the gravitational instability of modes
whose physical wavelength exceeds a critical value (Jeans wavelength)
\citep{Khlopov_Malomed_Zeldovich_1985,Nambu_Sasaki_1990,Fabris_Martin_1997}.
That critical value can be obtained by expanding the potential at
its minimum.
The sign of the quartic term determines whether the gravitational
instability is partially damped (positive sign as in (\ref{Eq:Pot_cosh}))
or increased (negative sign as in (\ref{Eq:Pot_cos})) relative to
the purely quadratic potential \citep{Khlopov_Malomed_Zeldovich_1985}.
This leads to additional power for $l \lesssim 10$ for
the cosine potential (\ref{Eq:Pot_cos}) in comparison to
the hyperbolic-cosine potential (\ref{Eq:Pot_cosh}).
This instability rules out potentials (\ref{Eq:Pot_cos}) and
(\ref{Eq:Pot_cosh}) with a parameter $\alpha \gtrsim 0.0065$
which leads to too much power at small values of $l$.
Such models are possible, however, in compact hyperbolic universes,
such that the Jeans wavelength is larger than the largest
wavelength of the eigenmode spectrum with respect to the Laplace-Beltrami
operator.
The largest wavelength belongs to the smallest eigenvalue of
the compact fundamental cell.
If the volume of the fundamental cell is small enough,
the smallest eigenvalue would be sufficiently large, such that
the largest wavelength would be smaller than the Jeans wavelength
and no instability would occur.
The effect of a finite size for nearly flat universes
with respect to low multipoles $l$ is discussed in
\citet{Aurich_Steiner_2000} in connection with a special
dark energy component having a constant $w_\phi=-2/3$.

\section{Monotonic Potentials}

\begin{figure}
\vspace*{-10pt}
\begin{center}
\hspace*{-25pt}\includegraphics[width=9cm]{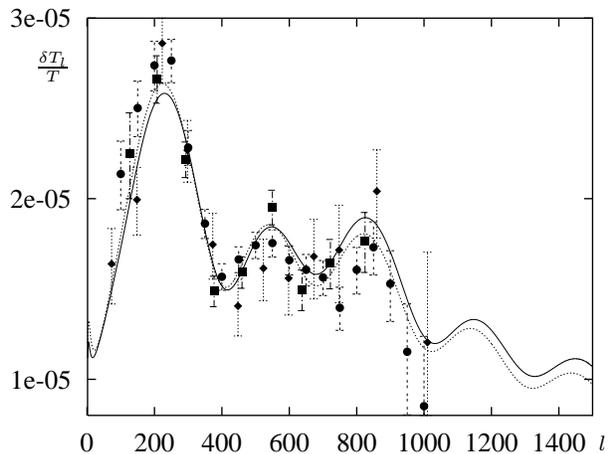}
\put(-15,7){$l$}
\put(-228,150){$\frac{\delta T_l}T$}
\end{center}
\vspace*{-15pt}
\caption{\label{Fig:Cl_exp}
The CMB angular power spectrum $\frac{\delta T_l}T$ is shown for
$\Omega_{\text{tot}} = 0.85$ for the exponential potential
(\protect\ref{Eq:potential_exp}) with $\alpha=0.0045$ (solid curve).
The dotted curve belongs to a model with $\alpha=0.006$ where a reduced
$\Omega_{\text{cdm}} = 0.25$ is used and
$\Omega_\phi$ is correspondingly increased to 0.55.
}
\end{figure}

\begin{figure}
\vspace*{-10pt}
\begin{center}
\hspace*{-10pt}\includegraphics[width=9cm]{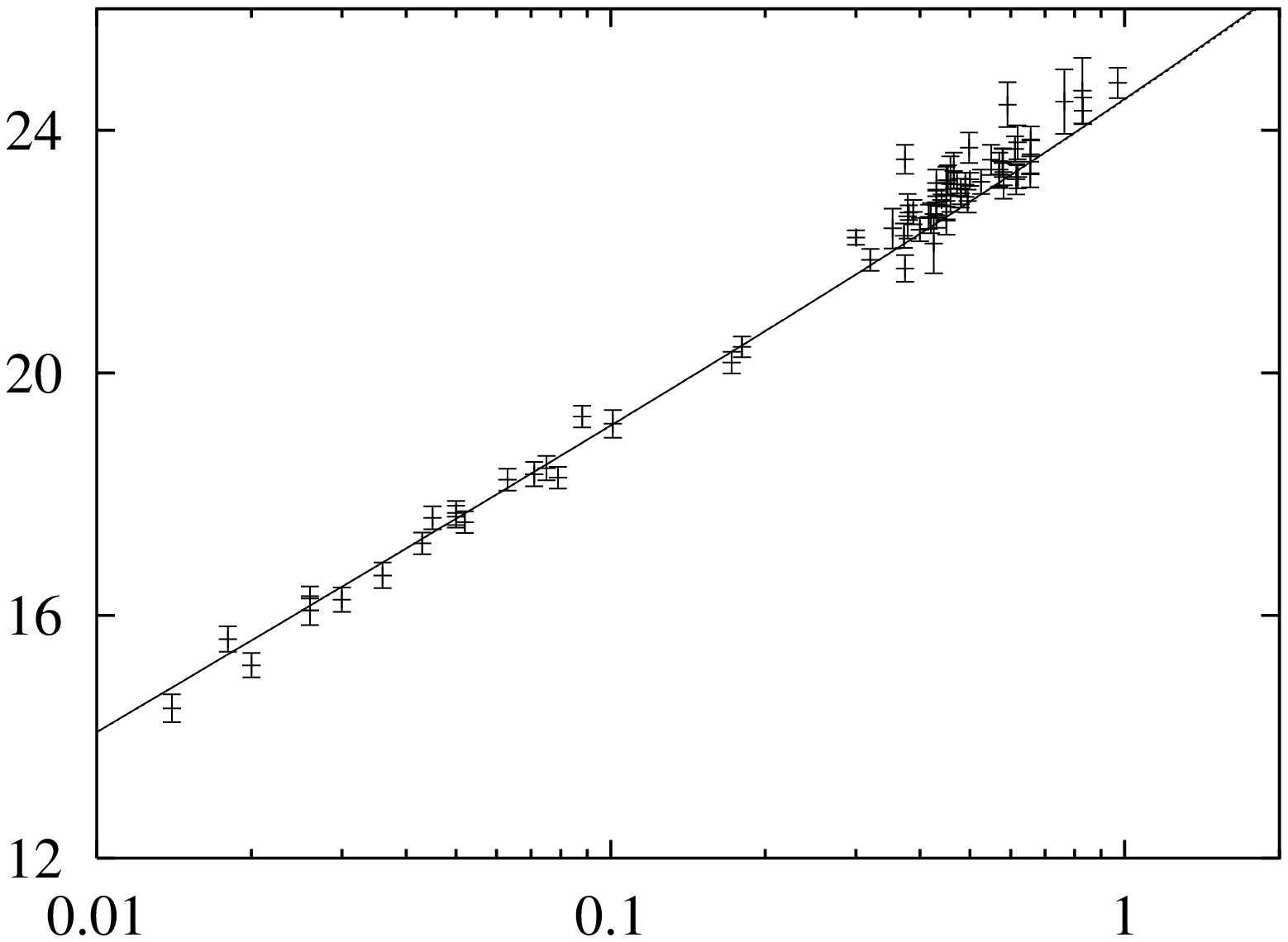}
\put(-25,7){$z$}
\put(-242,162){$m_{\text{B}}$}
\end{center}
\vspace*{-15pt}
\caption{\label{Fig:SN_exp}
The magnitude-redshift relation is shown for the same models as in figure
\ref{Fig:Cl_exp} in comparison with the supernovae Ia data.
}
\end{figure}

\begin{figure}
\begin{center}
\hspace*{-25pt}\includegraphics[width=9cm]{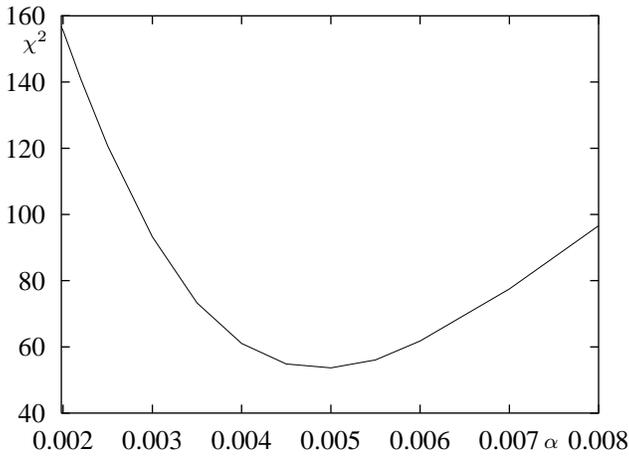}
\put(-38,7){$\alpha$}
\put(-235,158){$\chi^2$}
\end{center}
\vspace*{-15pt}
\caption{\label{Fig:Cl_exp_fit}
The $\chi^2$ values of the fit to 41 data points of the BOOMERanG,
MAXIMA-1 and DASI experiments for the potential (\ref{Eq:potential_exp})
with dependence on the parameter $\alpha$.
The parameter $\beta$ is chosen such that $\Omega_\phi(\eta_0)=0.45$,
i.\,e.\ $\Omega_{\text{tot}} = 0.85$.
}
\end{figure}

Models with monotonically decreasing potentials begin with small values of
the scalar field $\phi$ which then slides down the potential to
ever increasing values of $\phi$.
The potentials with a minimum discussed above show
that the location of the first acoustic peak depends on
$\alpha$ for fixed $\Omega_\phi(\eta_0)$.
Such behavior is also found in the case of the exponential
potential (\ref{Eq:potential_exp}).
The inverse power potential (\ref{Eq:potential_power}), however,
displays a nearly constant location of the first peak by
specifying $\Omega_\phi(\eta_0)$ independent of $\gamma$
in the range $3\leq \gamma \leq 7$.
E.\,g., for $\Omega_\phi(\eta_0)=0.5$, i.\,e.\ $\Omega_{\text{tot}}=0.9$,
the first peak lies at $l\simeq 225$.
Thus the exponential potential is more flexible with respect to the
peak position and, hence, does a better job in describing the observations.
An example for $\Omega_\phi=0.45$, i.\,e.\ for $\Omega_{\text{tot}} = 0.85$,
with $\alpha=0.0045$ is shown in figures \ref{Fig:Cl_exp} and \ref{Fig:SN_exp},
where the angular power spectrum and the magnitude-redshift relation
is shown, respectively.
These figures also show a model with $\alpha=0.006$
where the quintessence contribution is increased to $\Omega_\phi=0.55$
at the expense of the cold dark matter contribution
which is lowered to $\Omega_{\text{cdm}} = 0.25$.
The first peak lies at $l=230$ for $\Omega_{\text{cdm}}=0.35$,
and at $l=224$ for $\Omega_{\text{cdm}}=0.25$.
The second and third acoustic peaks are also at the correct positions.
Both models are compatible with the current CMB
($\chi^2 = 54.9$ for $\Omega_{\text{cdm}}=0.35$ and
$\chi^2 = 38.3$ for $\Omega_{\text{cdm}}=0.25$)
and supernovae data.
The low $\Omega_{\text{cdm}}$ model provides the best match to
the CMB data but it gives a power spectrum $P(k)$ of the LSS
which is too low as will be discussed in the next section.
Both models give an age of 11.2 Gyr being at the low end of the allowed range.
In these models $w_\phi$ has not yet reached its asymptotic value zero.
In the recent past, which is tested by the supernovae data,
$w_\phi$ had values of order $-0.1$.
Figure \ref{Fig:Cl_exp_fit} shows the dependence of $\chi^2$ on the parameter
$\alpha$ for $\Omega_\phi=0.45$, i.\,e.\ $\Omega_{\text{tot}}=0.85$.

%%%%%%%%%%%%%%%%%%%%%%%%%%%%%%%%%%%%%%%%%%%%%%%%%%%%%%%%%%%%%%%%%%%%%%%%%%%%%
%%%%%%%%%%%%%%%%%%%%%%%%%%%%%%%%%%%%%%%%%%%%%%%%%%%%%%%%%%%%%%%%%%%%%%%%%%%%%

%%%%%%%%%%%%%%%%%%%%%%%%%%%%%%%%%%%%%%%%%%%%%%%%%%%%%%%%%%%%%%%%%%%%%%%%%%%%%
%%%%%%%%%%%%%%%%%%%%%%%%%%%%%%%%%%%%%%%%%%%%%%%%%%%%%%%%%%%%%%%%%%%%%%%%%%%%%

%%%%%%%%%%%%%%%%%%%%%%%%%%%%%%%%%%%%%%%%%%%%%%%%%%%%%%%%%%%%%%%%%%%%%%%%%%%%%
%%%%%%%%%%%%%%%%%%%%%%%%%%%%%%%%%%%%%%%%%%%%%%%%%%%%%%%%%%%%%%%%%%%%%%%%%%%%%

Another class of quintessence models specifies the
equation of state $w_\phi$ instead of giving the potential.
It is even possible to construct potentials which lead to
a constant $w_\phi$.
The form of such potentials depends on the other energy components of the
model.
The simplest cases are a pure quintessence model,
which leads to the exponential potential (\ref{Eq:potential_exp}),
and a radiation-dominated model, which gives the inverse power potential
(\ref{Eq:potential_power}).
With more components, analytic expressions of the potentials can be given
only for special values of $w_\phi$.
For $w_\phi=-\frac 13$, the potential for a three-component model  
with radiation, matter and quintessence reads
\citep{Aurich_Steiner_2002}
\begin{equation}
\label{Eq:potential_w_ein_drittel}
V(\phi) \; = \; \frac{A}{\left(
\tilde \eta \sinh\left(B\phi\right) +
\cosh\left(B\phi\right) - 1\right)^2}
\end{equation}
with
$$
A = \frac 83 \, \frac{\Omega_\phi \tilde \Omega_2^2}{\Omega_{\text{m}}^2}
\, \varepsilon_{\text{crit}}
\hspace{6pt} , \hspace{6pt}
B = \frac{2\sqrt \pi}{m_p} \, \sqrt{\frac{\tilde \Omega_2}{\Omega_\phi}} \,
\hspace{6pt} , \hspace{6pt}
\tilde \eta = 2 \, \frac{\sqrt{\tilde \Omega_2 \Omega_{\text{rad}}}}
{\Omega_{\text{m}}}
$$
and $\tilde \Omega_2 = 1 - \Omega_{\text{rad}} - \Omega_{\text{m}}$.
The initial condition must be $\phi_{\text{in}}=0$ in order to
enforce a constant $w_\phi$ also at the earliest times.
This potential interpolates between the inverse power potential during the
radiation epoch and the exponential potential during the
quintessence epoch.
The parameters $A$, $B$ and $\tilde \eta$ are completely determined by
the cosmological parameters at $\eta_0$.
The CMB angular power spectrum is shown in figure \ref{Fig:Cl_w13}
for the two cases $\Omega_\phi = 0.5$ and $\Omega_\phi = 0.45$,
i.\,e.\ $\Omega_{\text{tot}} = 0.9$ and $\Omega_{\text{tot}} = 0.85$,
respectively.
The first peak occurs at $l=222$ and $l=234$, respectively.
The $\Omega_{\text{tot}} = 0.85$ model describes the data slightly
better than the model with $\Omega_{\text{tot}} = 0.90$,
and the second and third peaks also match better
($\chi^2 = 58.3$ for $\Omega_\phi=0.50$ and
$\chi^2 = 54.0$ for $\Omega_\phi=0.45$).
The magnitude-redshift relation is nearly identical for both models,
such that both curves cannot be separated in figure \ref{Fig:SN_w13}.
The age of the universe is 11.7 Gyr in both cases.

\begin{figure}
\vspace*{-10pt}
\begin{center}
\hspace*{-25pt}\includegraphics[width=9cm]{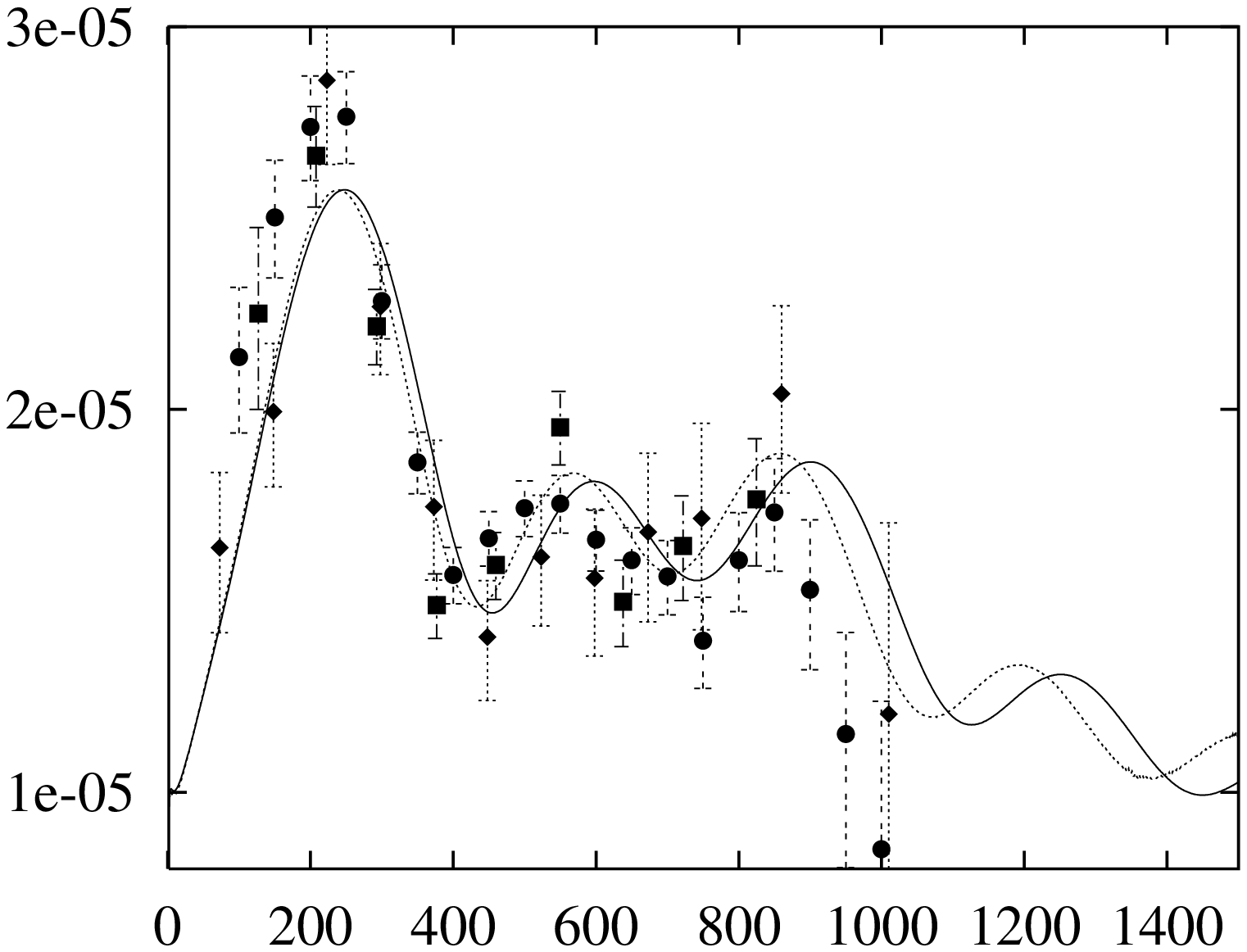}
\put(-15,7){$l$}
\put(-228,150){$\frac{\delta T_l}T$}
\end{center}
\vspace*{-15pt}
\caption{\label{Fig:Cl_w23}
The CMB angular power spectrum $\frac{\delta T_l}T$ is shown for
the potential (\ref{Eq:potential_w_zwei_drittel}) leading to
$w_\phi = -\frac 23$.
The solid curve belongs to $\Omega_\phi = 0.45$,
i.\,e.\ $\Omega_{\text{tot}} = 0.85$, and the
dotted curve to $\Omega_\phi = 0.5$, i.\,e.\ $\Omega_{\text{tot}} = 0.9$.
}
\end{figure}

\begin{figure}
\vspace*{-10pt}
\begin{center}
\hspace*{-10pt}\includegraphics[width=9cm]{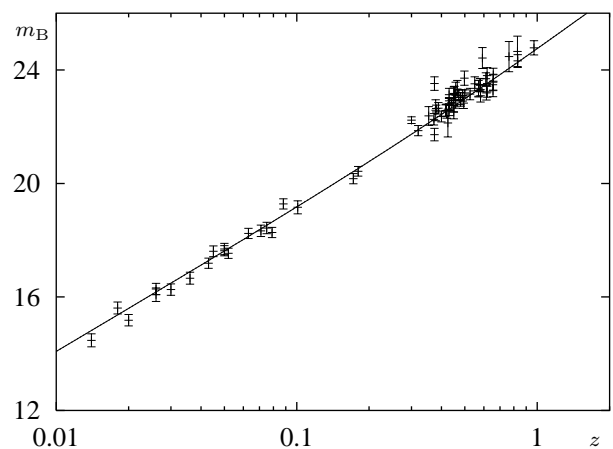}
\put(-25,7){$z$}
\put(-242,162){$m_{\text{B}}$}
\end{center}
\vspace*{-15pt}
\caption{\label{Fig:SN_w23}
The magnitude-redshift relation is shown for the same two models as in figure
\ref{Fig:Cl_w23} in comparison with the supernovae Ia data.
}
\end{figure}

Another potential can be derived 
for a two-component model consisting of matter and quintessence
with $w_\phi=-\frac 23$
\citep{Aurich_Steiner_2002}
\begin{equation}
\label{Eq:potential_w_zwei_drittel}
V(\phi) \; = \; \frac{A}{
\tilde \eta \sinh\left(B\phi\right) + \cosh\left(B\phi\right) - 1}
\end{equation}
with
$$
A = \frac 53 \, \frac{\Omega_\phi^2}{\Omega_2} \, \varepsilon_{\text{crit}}
\hspace{6pt} , \hspace{6pt}
B = \frac{\sqrt{8\pi}}{m_p}
\hspace{6pt} , \hspace{6pt}
\tilde \eta = 2 \, \frac{\sqrt{\Omega_{\text{m}} \Omega_\phi}}{\Omega_2}
$$
and $\Omega_2 = 1 - \Omega_{\text{m}} - \Omega_\phi$.
Here also the initial condition must be $\phi_{\text{in}}=0$.
Using this potential in a three-component model including radiation
gives an equation of state $w_\phi$ which deviates from $-\frac 23$
during the radiation epoch, but thereafter approaches very rapidly
the limiting value $-\frac 23$.
The corresponding models are shown in figures \ref{Fig:Cl_w23} and
\ref{Fig:SN_w23} for $\Omega_\phi = 0.5$ and $\Omega_\phi = 0.45$,
respectively.
It is seen in figure \ref{Fig:Cl_w23} that
the peaks in the CMB spectrum occur at values of $l$
which are larger than for the potential (\ref{Eq:potential_w_ein_drittel})
($\chi^2 = 72.1$ for $\Omega_\phi=0.50$ and
$\chi^2 = 111.3$ for $\Omega_\phi=0.45$).
The magnitude-redshift relation is, however, in excellent agreement
with the data as seen in figure \ref{Fig:SN_w23}.
These models give with 12.5 Gyr and 12.4 Gyr older universes than the
potential (\ref{Eq:potential_w_ein_drittel}) with $w_\phi=-\frac 13$.
Nevertheless, the model with $w_\phi=-\frac 13$ describes the CMB data
somewhat better than the model with $w_\phi=-\frac 23$.

\section{The Large-Scale Structure}

\begin{figure}
\vspace*{-10pt}
\begin{center}
\hspace*{-25pt}\includegraphics[width=9cm]{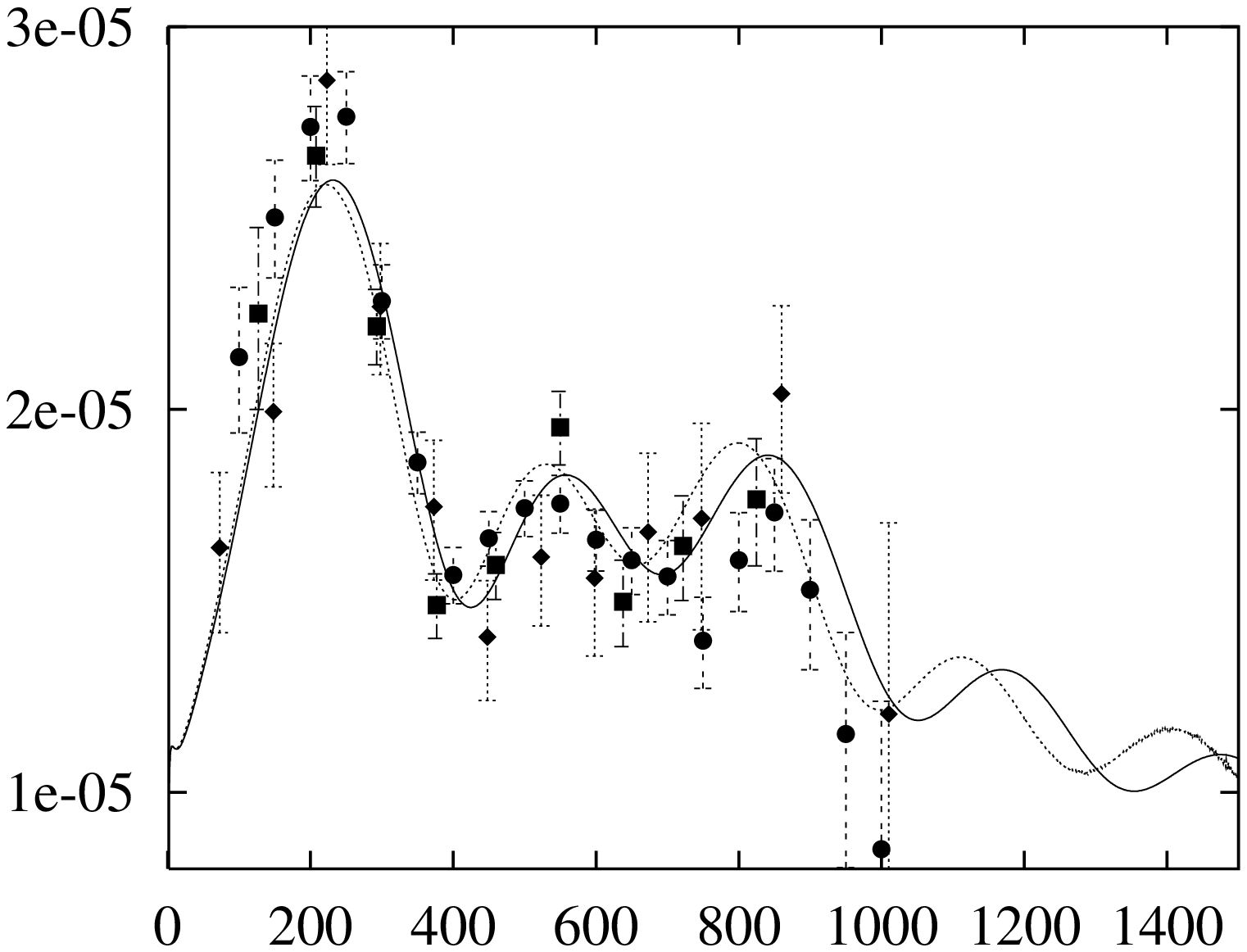}
\put(-15,7){$l$}
\put(-228,150){$\frac{\delta T_l}T$}
\end{center}
\vspace*{-15pt}
\caption{\label{Fig:Cl_w13}
The CMB angular power spectrum $\frac{\delta T_l}T$ is shown for
the potential (\ref{Eq:potential_w_ein_drittel}) with
$w_\phi = -\frac 13$.
The solid curve belongs to $\Omega_\phi = 0.45$,
i.\,e.\ $\Omega_{\text{tot}} = 0.85$, and the
dotted curve to $\Omega_\phi = 0.5$, i.\,e.\ $\Omega_{\text{tot}} = 0.9$.
}
\end{figure}

\begin{figure}
\vspace*{-10pt}
\begin{center}
\hspace*{-10pt}\includegraphics[width=9cm]{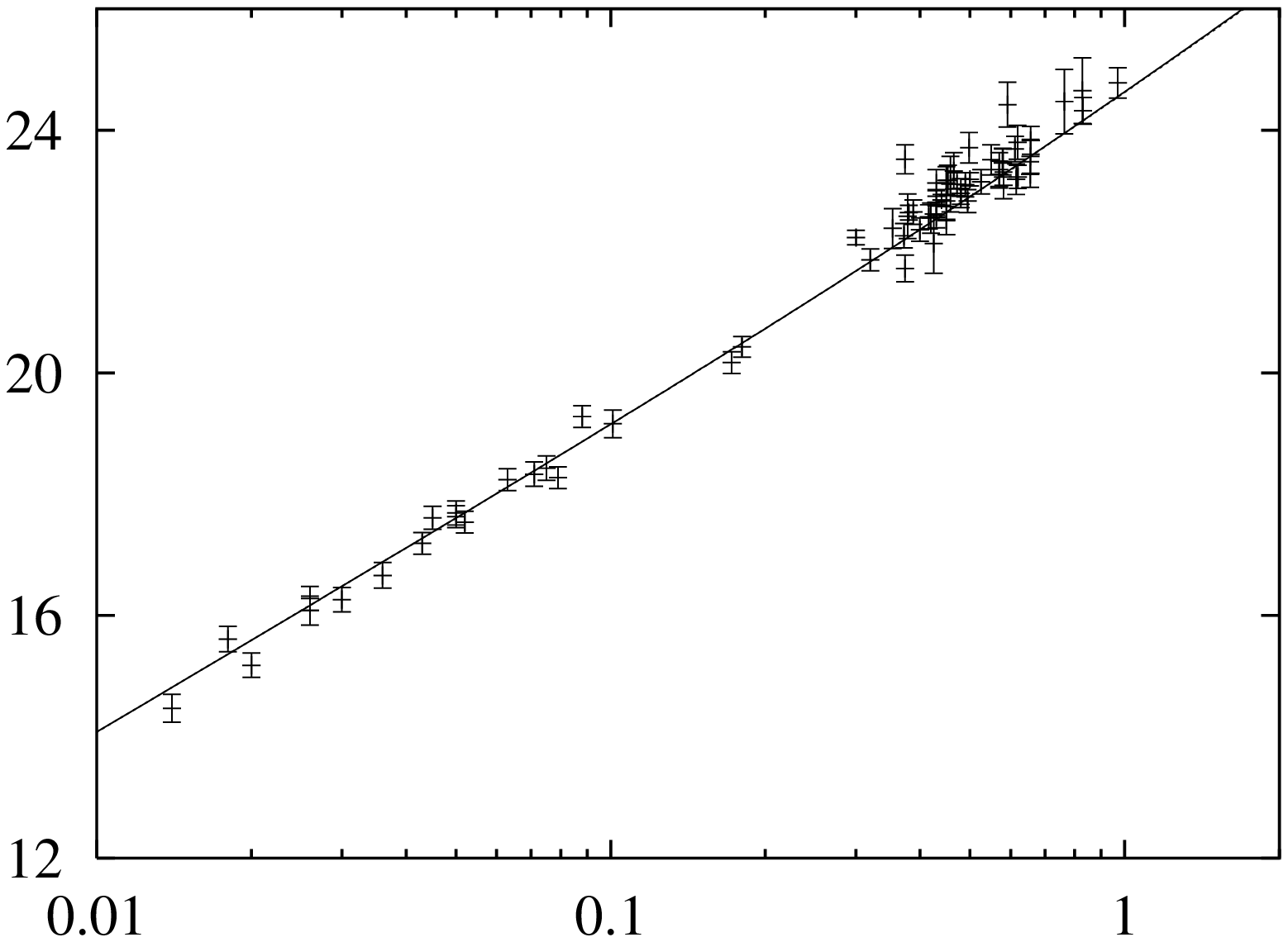}
\put(-25,7){$z$}
\put(-242,162){$m_{\text{B}}$}
\end{center}
\vspace*{-15pt}
\caption{\label{Fig:SN_w13}
The magnitude-redshift relation is shown for the same two models as in figure
\ref{Fig:Cl_w13} in comparison with the supernovae Ia data.
}
\end{figure}

\begin{figure}
\vspace*{-10pt}
\begin{center}
\hspace*{-18pt}\includegraphics[width=9cm]{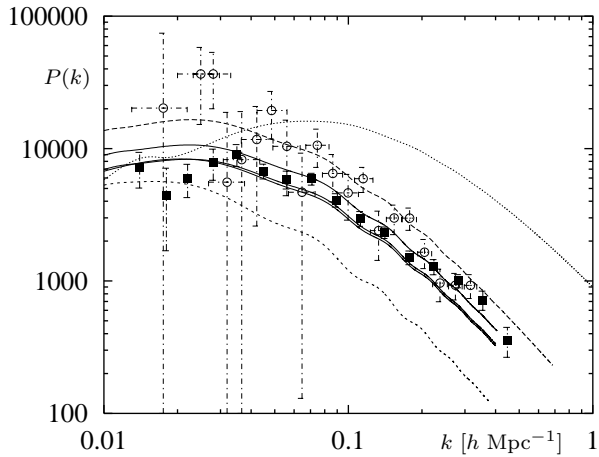}
\put(-75,7){$k$ [$h$ Mpc$^{-1}$]}
\put(-225,145){$P(k)$}
\end{center}
\vspace*{-15pt}
\caption{\label{Fig:P_k}
The LSS power spectra $P(k)$ $[h^{-3} \text{Mpc}^3]$ are shown as solid
curves for the three quintessence models which match the CMB and supernovae
data (the hyperbolic-cosine potential (\protect\ref{Eq:Pot_cosh})
with $\alpha=0.00375$,
the exponential potential (\protect\ref{Eq:potential_exp})
with $\alpha=0.0045$ and the potential (\ref{Eq:potential_w_ein_drittel})
with $\Omega_{\text{tot}}=0.85$).
$P(k)$ of the exponential model from figure \protect\ref{Fig:Cl_exp}
with a low cdm contribution $(\Omega_{\text{cdm}}=0.25, \Omega_\phi=0.55)$
is shown as a short dashed curve.
Also shown is a CDM model with $\Omega_{\text{cdm}}=0.93$ and
$\Omega_{\text{b}}=0.05$ (dotted curve) and a
$\Lambda$CDM model with $\Omega_{\text{cdm}}=0.35$, $\Omega_{\text{b}}=0.05$
and $\Omega_{\text{vac}}=0.58$ (long dashed curve).
The data points are from \protect\citet{Peacock_Dodds_1994}
(full squares) and from the IRAS PSC$z$
\protect\citep{Hamilton_Tegmark_Padmanabhan_2000} (circles).
}
\end{figure}

\begin{figure}
\vspace*{-10pt}
\begin{center}
\hspace*{-17pt}\includegraphics[width=9cm]{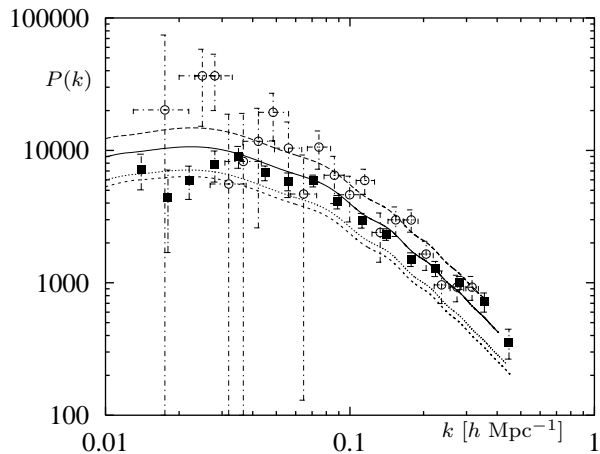}
\put(-75,12){$k$ [$h$ Mpc$^{-1}$]}
\put(-225,145){$P(k)$}
\end{center}
\vspace*{-15pt}
\caption{\label{Fig:P_k_cosh}
The LSS power spectra $P(k)$ $[h^{-3} \text{Mpc}^3]$ are shown for
the four quintessence models shown in figures \protect\ref{Fig:Cl_cosh}
to \protect\ref{Fig:w_cosh}, i.\,e.\ for the hyperbolic-cosine potential
(\protect\ref{Eq:Pot_cosh}) for $\Omega_{\text{tot}}=0.9$
with $\alpha=0.00237$ (long dashed curve), $\alpha=0.00375$ (solid curve),
$\alpha=0.0056$ (short dashed curve) and $\alpha=0.0058$ (dotted curve).
The data points are the same as in figure \protect\ref{Fig:P_k}.
}
\end{figure}

\begin{figure}
\vspace*{-10pt}
\begin{center}
\hspace*{-17pt}\includegraphics[width=9cm]{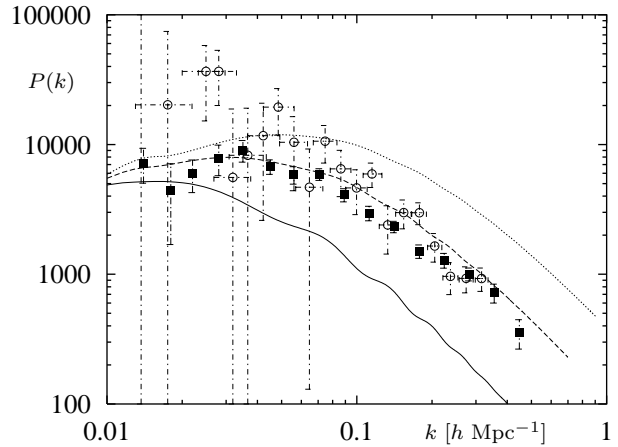}
\put(-75,12){$k$ [$h$ Mpc$^{-1}$]}
\put(-225,145){$P(k)$}
\end{center}
\vspace*{-15pt}
\caption{\label{Fig:P_k_exp_cdm}
The LSS power spectra $P(k)$ $[h^{-3} \text{Mpc}^3]$ are shown for
three quintessence models with the exponential potential
(\protect\ref{Eq:potential_exp}) using $\alpha=0.006$
for $\Omega_{\text{tot}}=0.9$
with $\Omega_{\text{cdm}}=0.25$ (solid curve),
$\Omega_{\text{cdm}}=0.5$ (dashed curve),
and $\Omega_{\text{cdm}}=0.75$ (dotted curve).
The data points are the same as in figure \protect\ref{Fig:P_k}.
}
\end{figure}

A further data set which the quintessence models have to explain
is the power spectrum $P(k)$ of the large-scale structure (LSS).
In the following we compare the models with the power spectrum $P(k)$
obtained from a compilation of galaxy cluster data by
\citet{Peacock_Dodds_1994} and with the decorrelated $P(k)$ by
\citet{Hamilton_Tegmark_Padmanabhan_2000}
using the IRAS Point Source Catalog Redshift Survey (PSC$z$)
\citep{Saunders_et_al_2000}.
These power spectra are shown in figures \ref{Fig:P_k} to \ref{Fig:P_k_exp_cdm}
as full squares \citep{Peacock_Dodds_1994} and as circles
\citep{Hamilton_Tegmark_Padmanabhan_2000}.
The linear regime is at wave numbers $k\lesssim 0.3 h {\text{Mpc}}^{-1}$,
where one expects a scale-independent bias parameter
$b = (P_{\text{Peacock-Dodds}}(k)/P(k))^{1/2}$.
For smaller scales, a scale-dependent bias is expected from
$N$-body simulations.
A nearly flat $\Lambda$CDM model (long dashed curve) is presented in 
figure \ref{Fig:P_k} together with a CDM model without dark energy
(dotted curve), where the latter has too much power at small scales.
In addition figure \ref{Fig:P_k} shows as solid curves the power spectra for
the three quintessence models that agree with the CMB and supernovae data:
the hyperbolic-cosine potential (\protect\ref{Eq:Pot_cosh})
with $\alpha=0.00375$ for $\Omega_{\text{tot}}=0.9$
(belonging to the highest lying of the three solid curves in
figure \ref{Fig:P_k}),
the exponential potential (\protect\ref{Eq:potential_exp})
with $\alpha=0.0045$ for $\Omega_{\text{tot}}=0.85$,
and the potential (\ref{Eq:potential_w_ein_drittel})
for $\Omega_{\text{tot}}=0.85$ with the CMB normalization
used in figures \ref{Fig:Cl_cosh}, \ref{Fig:Cl_exp} and
\ref{Fig:Cl_w13}, respectively.
All three spectra show a very similar behavior and
match the data obtained by Peacock and Dodds.
Thus no bias factor $b$ is required for these models.

%%%%%%%%%%%%%%%%%%%%%%%%%%%%%%%%%%%%%%%%%%%%%%%%%%%%%%%%%%%%%%%%%%%%%%%%%%%%%
%%%%%%%%%%%%%%%%%%%%%%%%%%%%%%%%%%%%%%%%%%%%%%%%%%%%%%%%%%%%%%%%%%%%%%%%%%%%%
%%%%%%%%%%%%%%%%%%%%%%%%%%%%%%%%%%%%%%%%%%%%%%%%%%%%%%%%%%%%%%%%%%%%%%%%%%%%%
%%%%%%%%%%%%%%%%%%%%%%%%%%%%%%%%%%%%%%%%%%%%%%%%%%%%%%%%%%%%%%%%%%%%%%%%%%%%%
%%%%%%%%%%%%%%%%%%%%%%%%%%%%%%%%%%%%%%%%%%%%%%%%%%%%%%%%%%%%%%%%%%%%%%%%%%%%%
%%%%%%%%%%%%%%%%%%%%%%%%%%%%%%%%%%%%%%%%%%%%%%%%%%%%%%%%%%%%%%%%%%%%%%%%%%%%%
%%%%%%%%%%%%%%%%%%%%%%%%%%%%%%%%%%%%%%%%%%%%%%%%%%%%%%%%%%%%%%%%%%%%%%%%%%%%%
%%%%%%%%%%%%%%%%%%%%%%%%%%%%%%%%%%%%%%%%%%%%%%%%%%%%%%%%%%%%%%%%%%%%%%%%%%%%%

It is well known that quintessence models have less power than
a $\Lambda$CDM model where the dark energy is provided by the
time-independent vacuum energy density.
This is due to the suppression of growth of perturbations
during the dark energy-dominated epoch.
This epoch begins generally earlier in quintessence models
than in $\Lambda$CDM models
(for a discussion of the flat case, see e.\,g.\
\citet{Coble_Dodelson_Frieman_1997,Wang_Steinhardt_1998,%
Skordis_Albrecht_2000,Doran_Schwindt_Wetterich_2001}).
To emphasize the point, figure \ref{Fig:P_k_cosh} shows $P(k)$
for the four hyperbolic-cosine quintessence models shown in figures
\protect\ref{Fig:Cl_cosh} to \protect\ref{Fig:w_cosh}.
These models differ in the duration of the dark energy-dominated epoch.
The long dashed curve in figure \ref{Fig:P_k_cosh} belonging to
$\alpha=0.00237$ shows the behavior for the frozen field,
where $w_\phi$ increases from initially $-1$ to only $-0.998$,
i.\,e.\ this case is practically indistinguishable
from a $\Lambda$CDM model with a constant vacuum energy density.
In this case the duration of the dark energy-dominated epoch is nearly
the same as for a $\Lambda$CDM model.
Increasing $\alpha$ from $\alpha=0.00375$ to $\alpha=0.0058$,
and thus enforcing a more dynamical field,
reduces the duration of the matter-dominated epoch and, in turn,
leads to lower LSS power as shown in figure \ref{Fig:P_k_cosh}.

%%%%%%%%%%%%%%%%%%%%%%%%%%%%%%%%%%%%%%%%%%%%%%%%%%%%%%%%%%%%%%%%%%%%%%%%%%%%%
%%%%%%%%%%%%%%%%%%%%%%%%%%%%%%%%%%%%%%%%%%%%%%%%%%%%%%%%%%%%%%%%%%%%%%%%%%%%%

%%%%%%%%%%%%%%%%%%%%%%%%%%%%%%%%%%%%%%%%%%%%%%%%%%%%%%%%%%%%%%%%%%%%%%%%%%%%%
%%%%%%%%%%%%%%%%%%%%%%%%%%%%%%%%%%%%%%%%%%%%%%%%%%%%%%%%%%%%%%%%%%%%%%%%%%%%%

The LSS power strongly depends on $\Omega_{\text{cdm}}$
since the cold dark matter perturbations are not coupled to
the radiation perturbations as is the case for the baryons
before recombination.
To stress that point,
figure \ref{Fig:P_k} shows also the results (short dashed curve) for
the exponential potential (\protect\ref{Eq:potential_exp})
again for $\Omega_{\text{tot}}=0.85$, where $\Omega_{\text{cdm}}$
is decreased to $\Omega_{\text{cdm}}=0.25$ and $\Omega_\phi$
increased to $\Omega_\phi=0.55$.
As shown in figure \ref{Fig:SN_exp} this modification alters
the magnitude-redshift relation only marginally.
We would like to emphasize that the low $\Omega_{\text{cdm}}$ model
gives the best description of the CMB data (see figure \ref{Fig:Cl_exp})
due to the enhanced amplitude of the first acoustic peak.
The LSS power, however, requires then a bias factor of $\sim 1.7$
for $k\gtrsim 0.04 h \text{Mpc}^{-1}$
as displayed by the short dashed curve in figure \ref{Fig:P_k}.
The LSS dependence on $\Omega_{\text{cdm}}$ is also demonstrated
in figure \ref{Fig:P_k_exp_cdm},
where the cold dark matter contribution is varied from
$\Omega_{\text{cdm}}=0.25$ to $\Omega_{\text{cdm}}=0.75$
for the exponential potential (\ref{Eq:potential_exp})
with $\Omega_{\text{tot}}=0.9$.
In comparison to $\Lambda$CDM models,
quintessence models need higher contributions of $\Omega_{\text{cdm}}$
in order to compensate for the shorter matter-dominated epoch.

The LSS power is not only reduced by replacing
the time-independent vacuum energy density by quintessence,
but is additionally reduced by the negative curvature 
in comparison to flat models.
The more negative the curvature, the more the LSS power is reduced
\citep{Wang_Steinhardt_1998}.
For the hyperbolic-cosine potential (\ref{Eq:Pot_cosh})
in the nearly frozen case $(w_\phi(\eta_0) < -0.97)$,
the LSS power increases by a factor of 1.2 by increasing
$\Omega_\phi$ from 0.45 to 0.58 in $\Omega_{\text{cdm}}=0.35$ models,
i.\,e.\ by increasing $\Omega_{\text{tot}}$ from 0.85 to 0.98,
despite the fact that the dark energy component is increased
which by itself reduces the LSS power.

%%%%%%%%%%%%%%%%%%%%%%%%%%%%%%%%%%%%%%%%%%%%%%%%%%%%%%%%%%%%%%%%%%%%%%%%%%%%%
%%%%%%%%%%%%%%%%%%%%%%%%%%%%%%%%%%%%%%%%%%%%%%%%%%%%%%%%%%%%%%%%%%%%%%%%%%%%%

\section{Conclusions}

We have studied quintessence models with various potentials
for a universe with negative curvature.
We assumed thereby $h=0.65$, $\Omega_{\text{b}}=0.05$
and a scale-invariant Harrison-Zel'dovich spectrum.
Several potentials have been found which are consistent with the current
CMB anisotropy, supernovae and LSS data as well with the age of the universe
with $\Omega_{\text{tot}}$ as small as $\Omega_{\text{tot}}=0.85$.
The best matches have been found for the hyperbolic-cosine potential
(\ref{Eq:Pot_cosh}), the exponential potential (\ref{Eq:potential_exp})
and the potential (\ref{Eq:potential_w_ein_drittel}) leading to a constant
$w_\phi=-\frac 13$.
In order to find agreement with the LSS data, it is necessary
to use $\Omega_{\text{cdm}}=0.35$, which together with
$\Omega_{\text{b}}=0.05$ (for $h=0.65$) gives
$\Omega_{\text{m}}=0.4$.
Dynamical mass determinations lead to values in the range
$\Omega_{\text{m}}=0.35 \pm 0.1 (2\sigma)$ \citep{Krauss_2001b}.
Therefore our value belongs to the higher values of the
allowed range.
However, such a relatively large value is necessary in order to
match the LSS observations.
If future observations will favor higher values of $\Omega_{\text{m}}$,
this would favor quintessence models in comparison to $\Lambda$CDM models,
because the latter have longer matter-dominated epochs and need
therefore lower values of $\Omega_{\text{cdm}}$ to achieve
the same LSS power, since $\Omega_{\text{b}}$ is fixed by BBN.

\subsection*{ACKNOWLEDGMENTS}
% \begin{acknowledgments}
We would like to thank the Scientific Supercomputer Centre (SSC) Karlsruhe
for the access to their computers.
% \end{acknowledgments}

\bibliography{../../bib_astro} 

\bibliographystyle{apalike}

\label{lastpage}

\end{document}